%% file: main.tex
\documentclass[journal]{vgtc}                     

\onlineid{1006}

\vgtccategory{Theoretical \& Empirical}

\title{Quantifying Visualization Vibes: \\ Measuring Socio-Indexicality at Scale}

\author{%
  \authororcid{Amy Rae Fox*}{0000-0003-0995-7899},
  \authororcid{Michelle Morgenstern*}{0000-0002-6321-6613}
  \authororcid{Graham M. Jones}{0000-0001-6435-7066},
  \authororcid{Arvind Satyanarayan}{0000-0001-5564-635X}
}

\authorfooter{
    \item
    \textbf{*} indicates equal contribution. Morgenstern \& Jones are affiliated with the MIT Department of Anthropology. Fox \& Satyanarayan are affiliated with MIT CSAIL. Email: [mjmorgen, amyfox, gmj, arvindsatya] @mit.edu

}
\abstract{What impressions might readers form with visualizations that go \textit{beyond} the data they encode? In this paper, we build on recent work that demonstrates the \textit{socio-indexical function} of visualization, showing that visualizations communicate more than the data they explicitly encode. Bridging this with prior work examining public discourse about visualizations, we contribute an analytic framework for describing inferences about an artifact's \textit{social provenance}.  Via a series of
attribution-elicitation
surveys, we offer descriptive evidence that these social inferences:
(1) can be studied asynchronously,
(2) are not unique to a particular sociocultural group or a function of limited data literacy,
and (3) may influence assessments of trust. Further, we demonstrate (4) how design features act in concert with the topic and underlying messages of an artifact's data to give rise to such `beyond-data' readings. We conclude by discussing the design and research implications of inferences about social provenance, and why we believe broadening the scope of research on human factors in visualization to include sociocultural phenomena can yield actionable design recommendations to address urgent challenges in public data communication.}

\keywords{semiotics, socio-indexicality, social provenance, engagement, visualization psychology, public data communication}

\teaser{
  \centering
  \includegraphics[width=1\linewidth, alt={This figure features two sets of data visualizations--each showing the original version of a visualization and an alternative version that has had its message obscured by replacing the titles and a labels with placeholder text. Below each set of images is a kernel density histogram showing a comparison between survey respondents' inferences regarding the design skill, intent, and political leaning of the visualization's maker when reading the message-obscured visualization versus the original version. For the first pair of stimuli, a correlation heatmap there was a moderate shift from professional to layperson and from intent to persuade to intent to inform with the message-obscured version being described as likely something from a "high school textbook" and the unobscured version as "just about recipes and spices", but inferences regarding the political leaning of the maker did not shift. In contrast, while both the message-obscured and obscured versions of the second set of stimuli, B1-C, provoked strong socio-indexical inferences, there was a dramatic shift in the identification of the maker's political leaning from right to left, going from being described as "an infographic on Fox News" to "a Democratic message" designed to capture the attention of conservatives. Taken together, the images in this figure demonstrate how inferences about the social provenance of a visualization (e.g. its maker and context of origin) arise can arise from design features, from data context, and from the interaction between the two.
}]{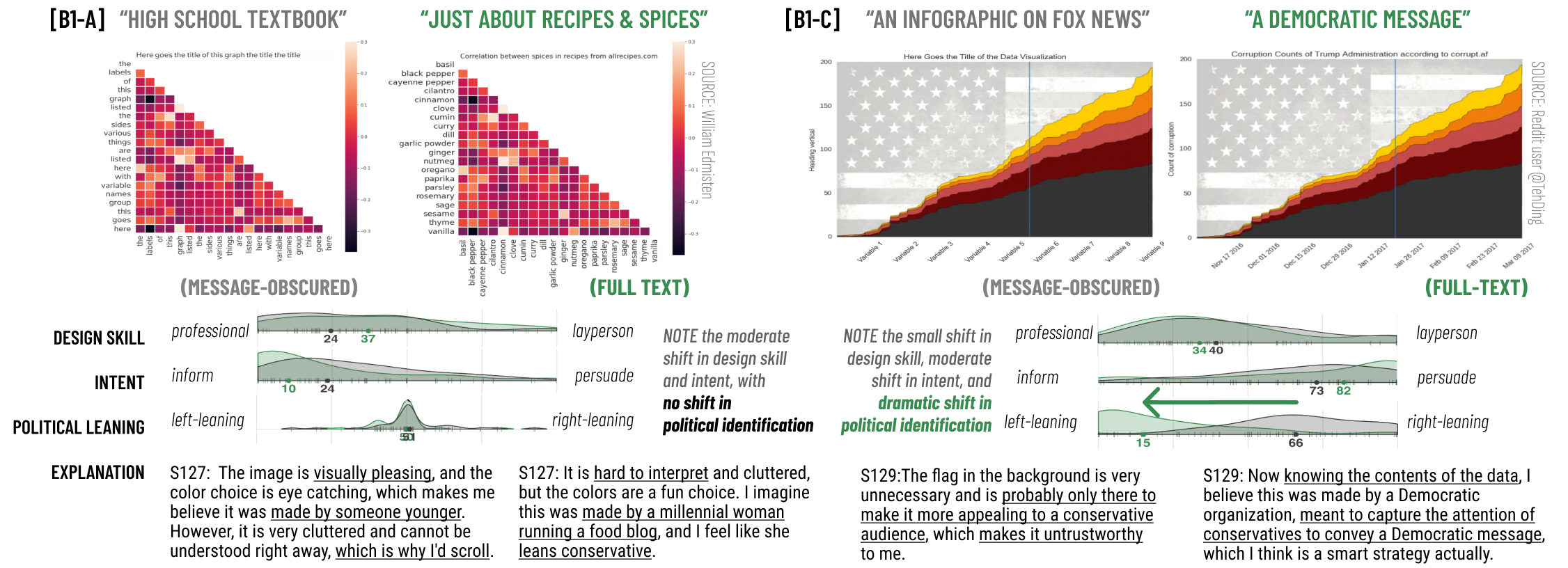}
  \caption{
  \textbf{Design \& Data Context.} Comparing shift in participant to message-obscured and original-text versions of stimuli from the Study 3 social attribution survey illustrates how  social inferences arise from the combination of design features and data context.}
  \label{fig:teaser}
}

\graphicspath{{FIGURES/}{./}} 

\usepackage{stfloats}

\usepackage{soul}

\usepackage[hang,flushmargin]{footmisc}

\usepackage{mathptmx}

\usepackage{censor}

\usepackage{lipsum}

\usepackage{etoolbox}

\usepackage{csquotes}

\usepackage{booktabs}
\usepackage{graphicx}
\usepackage{tabularx}
\usepackage{multirow}
\usepackage{makecell}
\usepackage{array}

\usepackage{tabu}

\usepackage{accessibility}
\usepackage{enumitem}

\newcommand{\code}[1]{\texttt{#1}}

\makeatletter
\AtBeginDocument{%
  \crefname{section}{§}{§§}
  \Crefname{section}{§}{§§}
  \crefformat{section}{\S\@gobble#1}
  \Crefformat{section}{\S\@gobble#1}
}
\makeatother

\begin{document}


\firstsection{Introduction}

\maketitle

\input{SECTIONS/01_introduction}
\input{SECTIONS/02_background}

\input{SECTIONS/04_methods}

\input{SECTIONS/05_results}

\input{SECTIONS/06_discussion}

\acknowledgments{
  This work was supported by MIT METEOR and PFPFEE fellowships for Amy Fox, and Amar G. Bose Fellowship, Alfred P. Sloan Fellowship and National Science Foundation award \#1900991.
}

\section*{Supplemental Materials}
\label{sec:supplemental_materials}
Supplemental materials are available on OSF at \url{https://doi.org/10.17605/OSF.IO/23HYX}.

\bibliographystyle{abbrv-doi-hyperref-narrow}
\bibliography{BIBS/main}
\end{document}

%% file: SECTIONS/01_introduction.tex
Although research has demonstrated that viewers have socially-situated and identity-driven responses to visualizations \cite{peckDataPersonalAttitudes2019a,heEnthusiasticGroundedAvoidant2024, Roth2003}, in offering guidelines for effective 
design, more attention has been paid to the relatively invariant human factors of graphical perception and graph comprehension \cite{franconeri2021science, Hegarty2011}. 
In a recent contribution we argue that visualization not only has a propositional function\textemdash conveying insights about depicted data\textemdash but 
\textit{also} a socio-indexical
function: conveying impressions of a visualization's social provenance (Morgenstern \& Fox et al. \cite{PAPER1}).
That is, visualizations can generate 
\textbf{socio-indexical inferences: }
\textbf{``inferences about a visualization’s social provenance based on socioculturally grounded attitudes toward the people, tools, and contexts with which specific visualization features are associated''}\cite{PAPER1}.
Visualizations convey impressions of the identities and characteristics of the actors
presumed to be involved in its production.
These impressions are evoked because the design features of a visualization 
can point toward\textemdash \textit{index}\textemdash social categories, contexts, and characteristics \footnote{ This is what distinguishes the function of socio-indexicality (as a semiotic mechanism) from its consequences, such as perceptions of trust. 
}.
In turn, these socio-indexical inferences can influence how people respond to or engage with a visualization in the first place.
The socio-indexical function of visualization is thus
critical to understanding how visualizations function as rhetorical objects in the world, and of clear relevance to visualization design.

The socio-indexical function of communication is a concept drawn from linguistic anthropology, 
which
has documented that listeners form impressions about speakers based on the formal features (e.g. accent, lexicon, etc.) of their speech \cite{dragojevicCenturyLanguageAttitudes2021, dragojevicLanguageAttitudes2017b, hall-lewSocialMeaningLinguistic2021}. For example, studies have documented that listeners hearing the same passages spoken with different accents often form different judgements about characteristics of the speakers, such as their intelligence, friendliness, and social status \cite{dragojevicLanguageAttitudes2017b, ladegaard2000language, aghaStereotypesRegistersHonorific1998a, eckert1999beltenhigh, kircherIntroductionLanguageAttitudes2022, dragojevicCenturyLanguageAttitudes2021}. 
Importantly, the impressions listeners form of speakers
need not be accurate in order to influence behaviour. 
For example, a classic experiment on language and social cooperation found that compliance with requests for theatre-goers
to complete questionnaires 
differed based on the dialect in which the request was delivered, 
and the participants' beliefs about the social groups who commonly use it
\cite{bourhis1976language}. 
Such work demonstrates that styles of speech serve as markers of a speaker's identity, and that listeners make social judgements about speakers' identities in relation to their own, which in turn shapes social interaction. 
This fundamentally semiotic phenomenon is not limited to natural language, and has been documented in forms of visual communication, including imagery~\cite{nakassisLinguisticAnthropologyImages2023}, typography~\cite{murphyFontroversyHowCare2017}, 
and graphic design~\cite{murphyFakeNewsWeb2023a}. 
Most recently, we offered evidence that data visualizations communicate socio-indexical meaning as well \cite{PAPER1}. 
Specifically, we demonstrated that readers made \textbf{social attributions: value-laden inferences about an artifact’s social provenance.}

The evidence provided in our initial
exploratory work was grounded in 
ethnographic methods and 
a purposive sociocultural sample  \cite{PAPER1}. 
In this paper, we 
provide a conceptual replication and extension using survey methods;
ultimately demonstrating that inferences about social provenance can be elicited asynchronously.
We further integrate our findings with two prior studies examining public discourse 
\cite{hullmanContentContextCritique2015,kauerPublicLifeData2021a} 
to synthesize an analytic framework\textemdash a typology\textemdash that can be used to construct instruments for eliciting social inferences or conducting content analysis of visualization discourse. 
We present the results of three attribution-elicitation studies.
First we replicate the phenomenon by engaging 
our prior population (n=78; users of the social media platform Tumblr), 
followed by a broader sample of US-based English-speakers (n=240; Prolific).  
After demonstrating that participants in both samples make inferences about social provenance, we test a hypothesis of considerable relevance to visualization design: that inferences about a visualization's makers can intervene in the relationship between its aesthetic-appeal and trustworthiness. 
Finally, in a third study where participants view message-obscured followed by non-obscured versions of visualizations (n=40; Prolific), 
we illustrate how readers make social inferences via a combination of a visualization's design features, topic and takeaway messages (see \autoref{fig:teaser}).  
Taken together, we provide converging evidence for the \textit{socio-indexical function} of visualization, demonstrating that
visualizations have the capacity to
communicate more than the semantic content of the data they encode.
Further, we offer quantitative evidence of how such inferences 
can impact readers' 
perceptions of
trust, and 
decisions about how 
a visualization should be engaged. 
We conclude by describing the implications of readings beyond data for research and design.
We discuss how 
social provenance may be intentionally or unintentionally encoded in 
a designer's work via the design decisions they make; 
and call for broadening the scope of human factors research in visualization, from centering readings of data through graphical perception and graph comprehension, to studies of socially-situated visualization behaviour. 

%% file: SECTIONS/02_background.tex
 


\section{Background and Related Work}
\label{background}

Kennedy \& colleagues argue that ``the quick extraction of accurate information is only one way of defining what constitutes 
an effective visualization'' \cite[pg. 17]{kennedyEngagingBigData2016}. 
They identify additional criteria, 
from the extent to which a visualization promotes learning, discourse, or persuasion, to evoking curiosity, surprise or empathy. 
To develop measures for this broader range of communicative objectives, 
researchers must extend their attention beyond graphical perception and cognition to a more general class of behaviour described as \textit{engagement}: ``the processes of looking, reading, interpreting and thinking that take place when people cast their eyes on data visualisations and try to make sense of them'' \cite[pg.2]{kennedyEngagingBigData2016}.
Here
we describe 
research addressing 
socially-situated 
behaviour, including 
engagement \cite{mahyarTaxonomyEvaluatingUser2015, kennedyFeelingNumbersEmotions2018,kennedyEngagingBigData2016, alebriEmbellishmentsRevisitedPerceptions2024}, 
reception\cite{heEnthusiasticGroundedAvoidant2024,peckDataPersonalAttitudes2019a} 
framing\cite{borkinMemorabilityVisualizationRecognition2016,kongFramesSlantsTitles2018,kongTrustRecallInformation2019} 
and most recently, social meaning\cite{PAPER1}.

\subsection{Sociocultural Context \& Engagement in Visualization}
\label{sub:background:engagement}

A growing number of studies are investigating engagement in the context of public data communication,
including 
analysis of online discourse \cite{hullmanContentContextCritique2015, kauerPublicLifeData2021a}, 
and in-person research with populations difficult to access online \cite{kennedyEngagingBigData2016,peckDataPersonalAttitudes2019a,heEnthusiasticGroundedAvoidant2024}.  
Analyzing comments made 
on The Economist magazine’s data journalism blog, Hullman \& colleagues observed that despite the blog's focus on data, discussions often centered on the broader context of a visualization rather than its content alone \cite{hullmanContentContextCritique2015}. 
Notably, critiques were frequently directed at the visualization’s \textit{creator}, reflecting an engagement with their design decisions and perceived intentions. 
In similar work, Kauer \& colleagues examined discourse in the visualization subreddit \code{r/dataisbeautiful} \cite{kauerPublicLifeData2021a}.
They developed a framework to categorize different forms of comments, focusing on the scope (e.g. subject) and form (e.g. genre) of public discourse around data. 
They found that users offered comments of various forms
(observations, hypotheses, clarifications, proposals) about various subjects 
(including the data, topic, and artifact design). 
In Section \ref{sec:methods:questions},  we extend this framework 
by including the form \emph{social attribution}, and expanding the range of subjects that can be characterized. 

Beyond online discourse, researchers have used interview methods to investigate how social factors shape individual differences in engagement with visualizations. Peck et al. conducted interviews with rural Pennsylvanians, eliciting explanations of how viewers assess the usefulness of different graphs \cite{peckDataPersonalAttitudes2019a}. The most salient theme emerging from their analysis was the extent to which a visualization's usefulness was grounded in assessment of its personal relevance, implicating a complex web of factors informing responses\textemdash from a participant’s political affiliation and educational background, to beliefs and values grounded in their sociocultural identities.  Similarly, He \& colleagues explored engagement 
by interviewing users of public spaces in Canada \cite{heEnthusiasticGroundedAvoidant2024}. They introduced the construct  \textit{information receptivity} to describe an individual's differential openness to external representations of information, and interpretations of the information.  Importantly, they defined receptivity as a transient state rather than a fixed trait: different stimuli and situations can provoke different reactions from the same individual.
We similarly view an individual's response to a graphical artifact as inherently shaped by their sociocultural context. In this work, we zoom in one level of abstraction, aiming to link a participant's reaction to a visualization with their 
\textit{socio-indexical} response to specific design features. If these features function analogously to indexicality in verbal and visual language, we anticipate they will influence both engagement and interpretive behaviour.


\vspace{-1mm}
\subsection{Framing \& Bias: The Role of Text and Data-Topic}
\label{sub:background:framing}

A key issue in public data communication is the extent to which visualizations can be designed to minimize bias. Although both critical \cite{aielloInventorizingSituatingTransforming2020, kennedyEngagingBigData2016, rettbergWaysKnowingData2020} and semiotic \cite{peirce-hoopes, FoxHollan_VisPsych_researchprogramme2023,rothWhatMeaningMeaning2004a} approaches clarify that a representation can never be entirely without bias, it is helpful to understand what features of a visualization trigger well-known cognitive effects like confirmation bias \cite{dimaraTaskbasedTaxonomyCognitive2018}, as well as how viewers conceive of bias with respect to data, visualizations, and narrative interpretations.  
Media and communication scholars have studied the role that titles in particular play in directing readers' attention \cite{blomClickBaitForwardreference2015, kuikenEffectiveHeadlinesNewspaper2017, scaccoCuriosityEffectInformation2020} and framing interpretation \cite{tannenbaumEffectHeadlinesInterpretation1953, scaccoCuriosityEffectInformation2020, ifantidouNewspaperHeadlinesRelevance2023}.
Visualization researchers have found these effects to be true of charts as well \cite{borkinMemorabilityVisualizationRecognition2016,kongFramesSlantsTitles2018,kongTrustRecallInformation2019}.
In fact, when the content of a title is slanted or even contradictory to the visualization's depicted data, readers typically recall gist messages aligned with the \textit{title}, rather than the data \cite{kongFramesSlantsTitles2018,kongTrustRecallInformation2019}.  
Most relevant to the issue of social provenance, Kong \& colleagues demonstrate that readers can differentiate between the credibility of a title, a visualization, and its data: the more slanted a title is, the more likely readers are to identify the title as biased, while maintaining that the information is impartial 
\cite{kongTrustRecallInformation2019}. This echos He et al.'s argument that readers can have different receptions to information (vs) interpretation, where a title acts as an interpretation of the ostensibly impartial encoded data
\cite{heEnthusiasticGroundedAvoidant2024}. 
In our exploratory inquiry we presented participants with a combination of original and message-obscured stimuli to gauge the extent to which inferences about social provenance are derived from a  visualization's textual content versus design features \cite{PAPER1}. 
We found that participants could make detailed inferences based on aesthetic features and
\textit{also} made inferences about the 
kind of people who would make charts about particular
topics. 
These indexical associations often unfolded into 
further inferences about the politics and value alignment between the reader and imagined makers.  
In the present work, we extend these findings via a repeated-measures study where we 
first elicit social attributions from a message-obscured image. 
We then present viewers with the original image and ask which if any of the impressions have changed.  In \hyperref[sec:results:study3-shift]{\S4.3} we discuss these results and how aesthetics and data-topic interact to give rise to social inferences.

\subsection{Social Meaning and Beyond-Data Reading}
\label{sub:background:paper1}

In linguistic anthropology, \textit{socio-indexical meaning} is understood as ``the constellations of inferences that can be drawn about speakers based on how they talk'' \cite[pg.1]{beltramaschwarz2024personae}. 
In a recent study inspired by research in linguistic anthropology on the semiotics of pragmatics \cite{silverstein1976shifters, aghaStereotypesRegistersHonorific1998a, irvine2001style, keaneSemioticIdeology2018}, 
we provide evidence for the socio-indexical function of visualization (Morgenstern \& Fox et al. \cite{PAPER1}). 
This descriptive work reveals that, like natural language, visualization not only has a referential function, whereby it explicitly encodes propositional meaning (insights about data), but also has a \textit{socio-indexical function} whereby formal features of visualization (i.e. structural and aesthetic design elements, such as chart type, colour, 
and typography) point to\textemdash that is, index\textemdash the social contexts, categories, and characteristics with which they are associated. 
Drawn from C.S. Peirce's triadic distinction between three ways a sign can make meaning \cite{peirce-hoopes}, indexicality is representation based on real or imagined causal connection or co-occurrence. 
While visualization design draws heavily on the iconic and symbolic functions of Peircean semiotics, we offer some of the first\footnote{Schofield et.al. \cite{schofieldIndexicalityVisualizationExploring2013a} discuss how the semiotic index might challenge the sign-world dichotomy implied by the visualization pipeline \cite{CardCHI11999}, though our work specifically addresses the \textit{socio-}indexical function \cite{PAPER1}.}
empirical evidence for the presence and consequences of socio-indexical
readings in visualization. 
Through a series of semi-structured interviews, 
we document how, in response to open-ended prompts, readers regularly offered comments about a visualization's "vibes"\textemdash the social contexts, categories, and characteristics evoked by a visualization's design. 
From "Microsoft Office vibes" to a "Boomer conservative vibe", readers made social attributions that ranged from concrete identifications of particular actors or types of actors
(human and nonhuman) involved in a visualization's production, to elaborate descriptions of their characteristics. 
We emphasize that both seemingly straightforward \textit{identifications} and clearly subjective \textit{characterizations} can be equally value-laden. Depending on a reader's own social positioning and sociocultural context, simple identifications of age/generation, political party, or even preferred social media platform can imply a host of far-from-neutral qualities. Importantly, we document that part of visualization's socio-indexical function is its impact on behaviour. While our study relied on self-reported impressions,
we document how readers described their socio-indexical readings as motivating decisions about how to interact with visualizations, as well as their reception and disposition toward the makers. For example, one participant commented:
\textit{``I'd probably just move along [without reading] before it made me angry,''} 
and another stated, \textit{``I would automatically distrust that person’s opinion''} \cite{PAPER1}.

%% file: SECTIONS/04_methods.tex
\vspace{-1.5mm}
\section{Exploring Socio-Indexicality At Scale}
\label{sec:methods}

The 
social attributions 
reported in our prior work
were elicited during semi-structured interviews facilitated by an ethnographer \cite{PAPER1}. 
However, the demand characteristics of interviews
are such that participants may be driven to respond in ways they perceive 
to conform to the interviewer's expectations.
Thus, our first aim is to determine whether it is possible to elicit social attributions without a live interviewer. Failure to do so would provide evidence that the prior results may have been compelled, 
or that the phenomena is so counter to dominant social norms that it cannot be elicited without an explicit permission structure 
scaffolded by an interviewer. 
Survey instruments
can provide converging evidence for the existence of the phenomenon, insofar as participants respond asynchronously, without the encouragement of an interviewer.
Furthermore, respondents
have the opportunity to offer indications of confidence, as well as
critique or non-response, through free-response text.
As a direct-elicitation technique, surveys also have demand characteristics; however, survey respondents experience less social pressure than interviewees (see\cref{discussion:limitations}).
Moreover, the patterns of variance in numeric responses between participants and across stimuli\textemdash especially in concert with free-response explanations\textemdash can serve as indications of random or biased responding, or alternatively, of salient features of the phenomenon. In the present studies, we aim to extend the observations of our prior work, determining if inferences about social provenance can be studied asynchronously, and if so, characterize the patterns that emerge from quantifying such inferences. Specifically, we 
ask: 

\begin{description}
    \item[RQ1] Can social inferences be elicited asynchronously, from a broader demographic sample?  If so, what patterns of variance and invariance emerge?
    \vspace{-1mm}
    \item[RQ2]  Do any types of inference affect the relationship between assessments of a chart's aesthetic \textit{beauty} and \textit{trust}?
    \vspace{-1mm}
    \item[RQ3]  How does the availability of the main message (e.g. data topic \& framing) of a visualization affect social inferences?
    \vspace{-1mm}
\end{description}

\noindent To address these questions, we conducted three studies by constructing attribution-elicitation surveys. \textbf{Study 1}  
conceptually replicates \cite{PAPER1}, eliciting social attributions from users of the social media platform Tumblr using a survey instrument to determine if the prior findings were resultant from the demand characteristics of an interview protocol (n = 78; Tumblr; RQ1, RQ2).  
\textbf{Study 2} 
extends
these findings 
beyond the original socio-cultural group, using the same survey instrument with a broader demographic sample to determine if the prior findings were idiosyncratic to participants in the sociocultural milieu of Tumblr, a platform known for its high metasemiotic awareness\cite{michelle-diss} (n = 240; Prolific; RQ1, RQ2).  
\textbf{Study 3} 
explores
if social attributions are affected by a graph's \textit{text} in combination with its design features. We shortened the survey instrument to gather responses to each visualization twice: first, a message-obscured version followed by the original untreated image (n = 40; Prolific; RQ3).

\begin{figure*}[tb]
  \centering 
    \includegraphics[width=1\textwidth,
    alt={Part A of this figure illustrates the survey procedure, as described in section 3.3. It shows how, after instructions, participants were randomly assigned to one of six stimuli blocks, which are all depicted in part B of the figure. Each block (B1 to B6) includes four message-obscured stimuli that vary in their degree of aesthetic embellishment. Also depicted is the common stimulus (B0-D) seen by all participants.
}
    ]{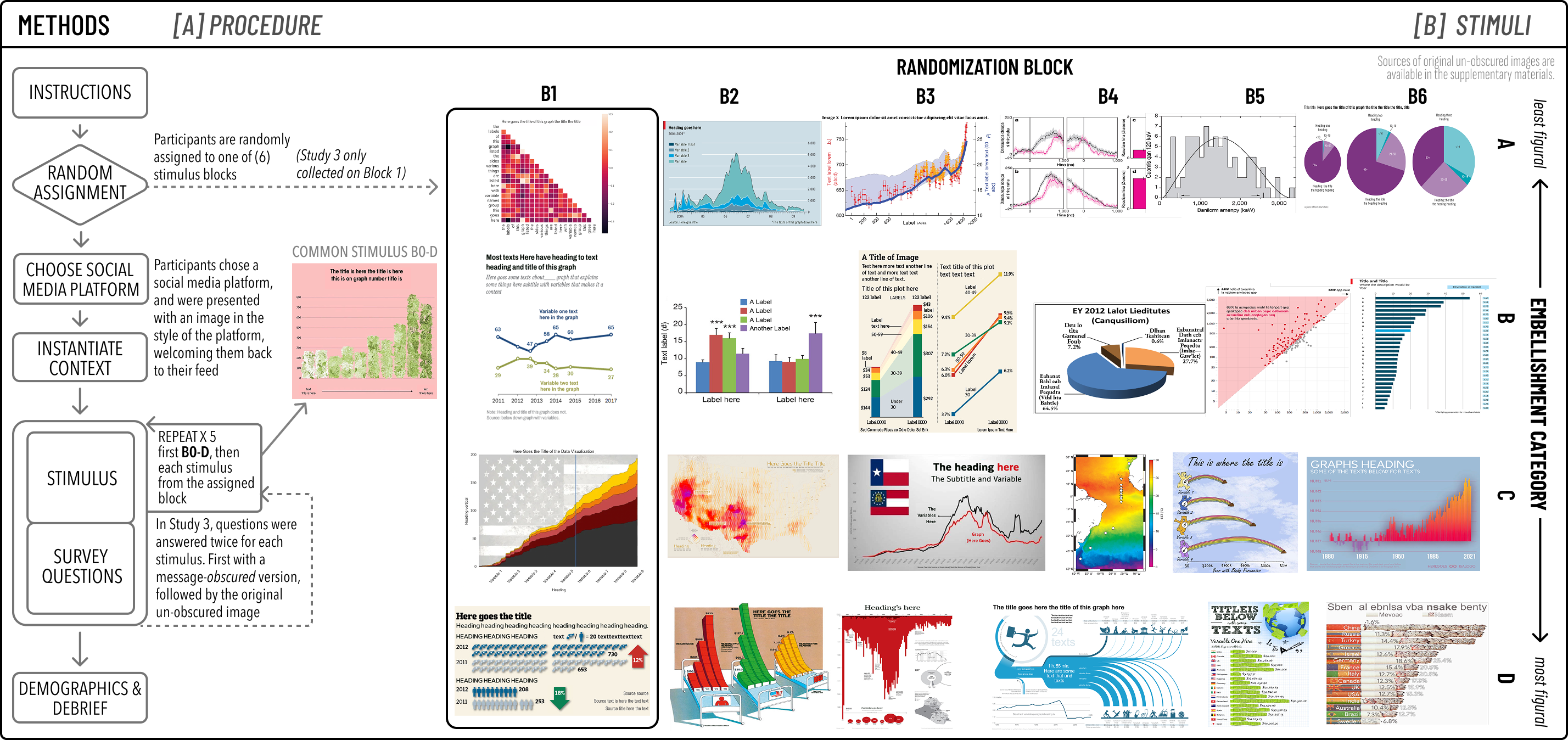}
  \caption{
  \textbf{Survey Procedure \& Stimuli}
  Part A—\textit{left} illustrates the structure of a participant's survey session.  Part B—\textit{right} depicts the full set of stimuli.
  }
  \label{fig:methods:diagram}
  \vspace{-5mm}
\end{figure*}
\subsection{Study Design}
\label{sec:methods:design}

Each study was implemented as a Qualtrics survey and framed to participants as ``an exploration of the salience of images on social media.'' Because our interaction with visualizations is fundamentally contextual, it is necessary to instantiate a specific situational context for engaging with stimuli\cite{hullmanRealWorldPeople2023a, Fox_VisPsych_theoriesmodels_2023}. We chose the grounding of a social media news feed because social media is a familiar context for encountering visualizations in the wild, and to afford comparison with data 
from our prior work \cite{PAPER1}. We used the framing of visual salience (rather than reading or interpretation) to reinforce the context of social media, and set the expectation that participants would be viewing images and offering impressions, rather than being evaluated on their graph reading skills. In each study, participants viewed a series of (v = 5) visualizations, and answered questions eliciting their impressions of each image.


\subsection{Participants}
\label{sec:methods:participants}

\textbf{Study 1} 
leveraged the sampling approach of \cite{PAPER1}, recruiting 
via blog post on the social media platform Tumblr, 
yielding ( n = 78 ) participants (36\% Female, 5\% Male, 40\% Non-binary / third gender, 17\% prefer to self-describe, 3\% prefer not to say) 
each compensated 
via 
\$10.00 gift card. 
\textbf{Study 2} used the same survey with a broader demographic sample, 
recruiting ( n = 240 ) US-based English-speaking adults via Prolific (54\% Female, 42\% Male, 3\% Non-binary, 1\% prefer not to say) 
each paid \$15.00/hour.
\textbf{Study 3} 
recruited ( n = 40 ) US-based English-speaking adults via Prolific (50\% Female, 47.5\% Male, 2.5\% Non-binary) 
paid \$15.00/hour.

\subsection{Procedure}
\label{sec:methods:procedure}

\cref{fig:methods:diagram}-A(\textit{left}) illustrates the survey procedure. To establish the situational context, 
participants 
selected
one of five social media platforms 
and were 
instructed to imagine 
engaging with that platform for the remainder of the survey. They then saw an image of a news feed from the platform, a pattern repeated between every trial to designate the transition to a new stimulus and reinforce the situational context of scrolling through a social media feed. Participants then proceeded to complete a sequence of five trials. In each, the stimulus was anchored to the top of the screen, with survey questions scrolling underneath such that the stimulus was always visible. Following the last trial, participants completed a short demographic questionnaire. In Studies 1\& 2, participants completed five trials (randomly assigned to one stimulus block) and response times ranged from $11$  to $227$ minutes, ($M = 45, SD = 26$).  In Study 3, participants completed five trials (stimulus Block 1), repeating each trial \textit{twice}: first viewing the message-obscured version of the stimulus, followed by the same set of questions while viewing the original (un-obscured) version. Response times 
ranged from $13$  to $110$ minutes, ($M = 41, SD = 20$).

\subsection{Materials: Visualization Stimuli}
\label{sec:methods:stimuli}

To explore the extent to which participants are able to make social attributions \textit{without} the context of data topic, we began by extending the corpus of stimuli used in \cite{PAPER1}. We first selected 
images that elicited particularly detailed indexical readings in the interviews, 
and then extended the corpus with visualizations from the MassVis dataset \cite{borkinWhatMakesVisualization2013}, optimizing for a variety of chart types and publishing sources. We developed \textit{message-obscured} versions of each image where all text was replaced with placeholders, mimicking typeface and spatial layout. To engage with prior work on aesthetics \cite{alebriEmbellishmentsRevisitedPerceptions2024, batemanUsefulJunkEffects2010}, we organized the images into four groups, positioned along a continuum of increasing aesthetic `embellishment', from \textit{most abstract} (group A: containing primarily marks as required by the chart type) to \textit{most figural} (category D: containing iconic imagery such as pictograms from which the data topic might reasonably be inferred). We selected one image that elicited consistently detailed
indexical responses in \cite{PAPER1} and designated that as a common stimulus to be seen by each participant (B0-D, see \autoref{fig:methods:diagram}). We organized the remaining stimuli into (6) randomization blocks containing (4) images, one from each embellishment group. In assigning stimuli to blocks we sought to maximize variance in chart type, publication source, and our own subjective assessment of visual style, such that no participant would see four stimuli that were similar along any one of those dimensions. \autoref{fig:methods:diagram}-B\textit{(right)} depicts the complete set of stimuli. 
Each participant in Studies 1 \& 2 was randomly assigned to one of the (6) blocks, and thus responded to questions about (5) images: the common stimulus (B0-D) and one image from each embellishment group in randomized order. In Study 3 we sought to determine how the context of data-topic affects social inferences, and elected to begin our exploration of this research question with data collection limited to one stimulus block. We chose Block 1 because three of the four stimuli include elements we believed may cause social inferences to change once the data topic was revealed. Thus each participant in Study 3 was assigned to Block 1, and answered questions about each image twice: first the \textit{message-obscured} version, followed by the original (un-obscured) visualization. 

\begin{figure}[!b]
  \centering 
      \vspace{-5mm}
\includegraphics[width=1\columnwidth,
alt={
This table features three columns illustrating our analytic framework for describing social provenance. The first column features categories of ACTOR, the subject or topic of the socio-indexical inference. These categories are: 1) MAKER (the entity responsible for an artifact, which is further subdivided into principal, the entity responsible for having the artifact made; author, the entity responsible for materially creating the artifact; and animator, the entity responsible for circulating the artifact), 2) MODE (the venue, source, or channel where an artifact is distributed, 3) TOOL (the instruments used in the artifact's production), 4) AUDIENCE (the individuals who receive and engage with the artifact), 5) DATA (the entities explicitly encoded in the artifact), and 6) ARTIFACT (the material--physical or digital--substance of the artifact itself). The second column of the table lists corresponding PROPERTIES, or aspects of the actor being described. These include properties such as age, gender, trustworthiness, occupation, medium, platform of circulation, specific tools, data literacy of the audience, or data topic. The third and final column features an example social attribution made about each listed category of actor drawn from the survey free responses.
}]{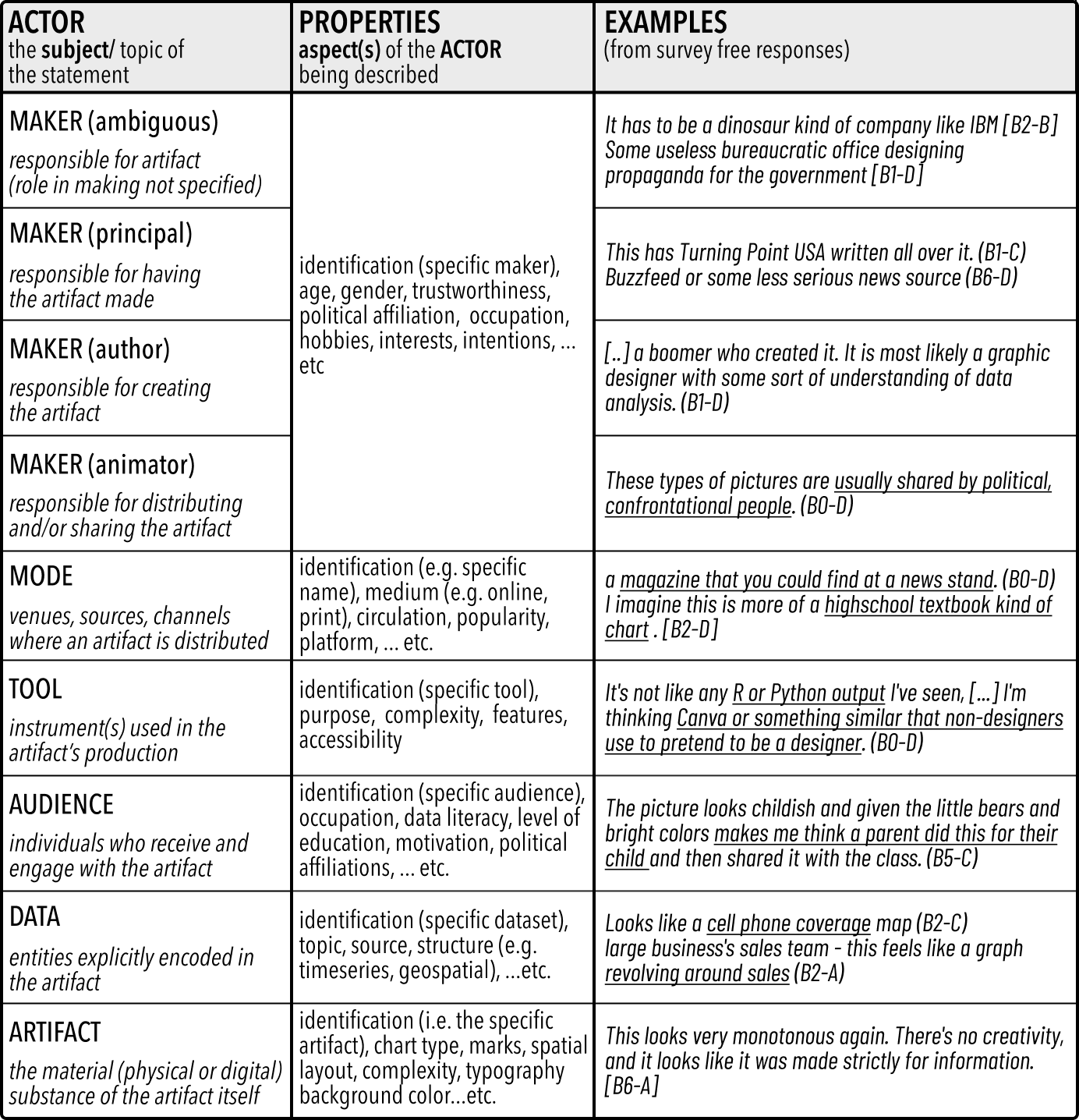}
  \caption{
    \textbf{Analytic Framework for Describing Social Provenance.} 
    Inferences about the social provenance of a visualization  are composed statements about the \textsc{properties} of \textsc{actors}, often expressed as chains of indexical associations, such as \textsc{actor(property)} → \textsc{actor(property) }
  }
  \label{fig:results:framework}
\end{figure}

\subsection{Materials: Analytic Framework \& Survey Questions}
\label{sec:methods:questions}

To aid in the development of survey questions, we developed an \textit{analytic framework} that can be used to describe the semantic content of discourse about social provenance. Kauer et al.'s  contribution analyzing discourse on the \code{r/ dataisbeautiful} subreddit served as the inspiration for this endeavour, and we began by evaluating the applicability of their framework, which identified the \textsc{Scope} and \textsc{type} of a comment made about a visualization; where \textsc{scope} characterizes the \textit{subject} of a comment, and \textsc{type} characterizes its \textit{form}\cite{kauerPublicLifeData2021a}. We chose to extend the \textsc{Scope} portion of this taxonomy to include the social \textsc{Actors} identified in our prior work \cite{PAPER1}.  
We reviewed the social attributions expressed in those interviews, and descended a level of detail to identify as many \textsc{Properties} of named \textsc{Actors} as possible. For example, interviewees described the  Maker's gender, age, occupation, interests, and political values, among others. We construe these as
\textsc{Properties} of the \textsc{maker}.
We added three additional \textsc{actors}: (1) \textsc{data}, (2) the visualization (\textsc{artifact}), and (3) \textsc{audience}; to better reflect the scope of inferences that might be made about a visualization's social origins and purpose. The resulting typology is illustrated in \autoref{fig:results:framework}
\footnote{See the \href{https://doi.org/10.17605/OSF.IO/23HYX}{supplemental materials} for a detailed translation between our structure and those offered by Kauer \& colleagues \cite{kauerPublicLifeData2021a} and Hullman \& colleagues\cite{hullmanContentContextCritique2015}.}
. To construct our survey instrument, we developed questions centered on the \textsc{Properties} 
we thought were most likely to yield actionable insights for visualization design; especially the viewer's beliefs about the identity, values, and competencies of the \textsc{Maker}.  The result was a combination of: (1) multiple choice questions (eliciting identifications), 
(2) semantic-differential scales \cite{heiseSemanticDifferentialAttitude1970} (eliciting characterizations), and (3) free-response text (eliciting explanations of the previous responses). Together these constituted the `long-form' survey used in Studies 1 \& 2. We developed a `short-form' version for Study 3 to accommodate the repeated-measures design without increasing 
participant fatigue. We removed questions about \textsc{Tools}, Behaviours, and a subset of questions about the \textsc{Maker} and \textsc{artifact} that were strongly correlated (e.g. \textsc{Artifact(Like)} was almost precisely correlated with \textsc{Artifact(Beauty)}). 
In the interest of space we illustrate the short-form survey questions in \autoref{fig:results:descriptives}, while the long-form is available in the \href{https://doi.org/10.17605/OSF.IO/23HYX}{supplemental materials}.

\subsection{Data Analysis}
\label{sec:methods:analysis}

Consistent with the descriptive nature of this work, we conducted an exploratory data analysis of quantitative survey measures, aiming to describe patterns of systematic variance and invariance across participants and stimuli. All statistical analyses were conducted in R ~\cite{r-lang}, and quantitative measures were standardized via z-score prior to modelling, though distributions are visualized in this paper using the original response scales for ease of reading. 
We also performed a thematic analysis of free response data, updating the analytic framework (\cref{fig:results:framework}) with additional \textsc{Actor(Properties)} that were present in respondents' explanations, and extracting representative quotations illustrating salient patterns in the quantitative results. 
\textit{Survey data, stimuli, model specifications and statistical tests are available alongside a reproducible analysis notebook in the \href{https://doi.org/10.17605/OSF.IO/23HYX}{supplemental materials}.}

%% file: SECTIONS/05_results.tex
\section{Results}
\label{sec:results}

In this section, we report the results of our attribution-elicitation studies organized thematically by research question. We begin by describing how the content of free-response text (\cref{sec:results:replication:free-response}), structure of variance in participant confidence scores (\cref{sec:results:replication:confidence}), and pattern of response to semantic-differential questions (\cref{sec:results:replication:semantic-differentials})  demonstrate that social attributions can be elicited asynchronously. 
We describe an exploratory factor analysis indicating a latent structure underlying the elicited social attributions (\cref{sec:results:replication:latent-structure}), and statistical models of the relationship between a viewer's assessment of a chart's beauty and the maker's trustworthiness, demonstrating how inferences of the maker's skill in data analysis, intent, and alignment with the viewer's own values moderate this relationship (\cref{sec:results-trust}). We conclude by describing which impressions of a visualization shift when formed with (vs) without the context of a graph's textual content (\cref{sec:results:study3-shift}). \textit{Quotations of free-response text are referenced by the survey ID and stimulus (e.g. \code{S101:B1-C}).}

\subsection{RQ1:Social Inferences Can be Elicited Asynchronously}
\label{sec:results-replication}

\subsubsection{Free-Response Explanations}
\label{sec:results:replication:free-response}

\begin{figure*}[p!]
\vspace{-5mm}
  \centering 
    \includegraphics[width=1 \linewidth,
    alt={This full page figure consists of a column displaying the questions and response options in the attribution elicitation survey, and five columns that display the distribution of responses for each study for each stimulus in block 1.  The questions on the survey are as follows. First are a series of semantic differential questions, implemented as sliders along a continuous scale of 0 to 100, where the 0 corresponds to the first adjective and 100 the second adjective.  Maker design skill: Are they more likely professional or layperson in graphic design?  Data Skill:  are they likely more professional or layperson in data analysis? Maker-Politics: are their politics more likely left leaning or right leaning? Maker-Alignment: do they likely share your values or do not share your values? Maker-Intention: is their intent more likely to inform or to persuade? Maker-Trust:  are they more likely untrustworthy or trustworthy? Chart-Beauty:  how aesthetically pleasing is the image (not at all to very much)?  The final four questions are single option multiple-choice. Maker-identity: who do you think is most responsible for having this image created? (options: an individual, organization, education, business, news, political) Maker-Age: what generation are they most likely from? (options: boomer, Gen X, millennial, Gen Z)  Maker-Gender:  what gender do they most likely identify with? (options: other, female, male). Behavior-Encounter: As you're scrolling through your feed if you see this image, what would you do? (options: scroll, or stop and read).  The visualizations next to each question represent the distribution of values for each stimulus for each question, and illustrate salient patterns that are discussed in the text of the manuscript.  Here we include the text annotations.  For maker design skill we see that images of complex chart tapes and scaffolding such as subtitles and illustrations were often attributed to makers of more professional design skills, while images that included tool defaults or bespoke elements for were more often attributed to makers with less skill in design. Attributions for Maker-Data-Skill were often associated with the complexity of the chart tape or complexity of the underlying data or statistical analysis, while simple chart types — especially bar charts — were often associated with makers with less skill in data analysis.  For the political leaning question there is a distinct difference in the pattern of responses for more abstract images where responses are clustered around 50\% – uncertain as to the political leaning of the maker – compared with the more embellished images where there were wider distributions with readers making some inference about the political affiliations that differed greatly between subjects.  Responses for the Maker-Alignment question tended to reflect correspondence with the ratings of the maker’s political values in the context of the viewers own political beliefs. Ratings for the Maker-Intention question largely corresponded with the categorical Maker-Identity question, where political and business makers were more likely to create images meant to persuade, and educational and news outlets were more likely intended to inform.  Inferences about the Maker-Identity varied widely between stimuli, largely with explanations that mentioned a combination of chart type and aesthetic features. For Maker-Gender we observe a stark difference in inferences about the gender of the stimulus B1-C which was largely attributed to males, and described as aggressive as it had a dark red color palette with an American flag background. In contrast is stimulus B0-D which was largely attributed to females, described as having a pink background and images of small plants.  Finally, for the Behavior-Encounter question,  thematic analysis of free response explanations for whether the individual would stop and read or scroll past the image indicated strong individual differences in how participants share or engage with content on the social media platform of choice in general, rather than specific features of the visualization driving their decisions.  
}]{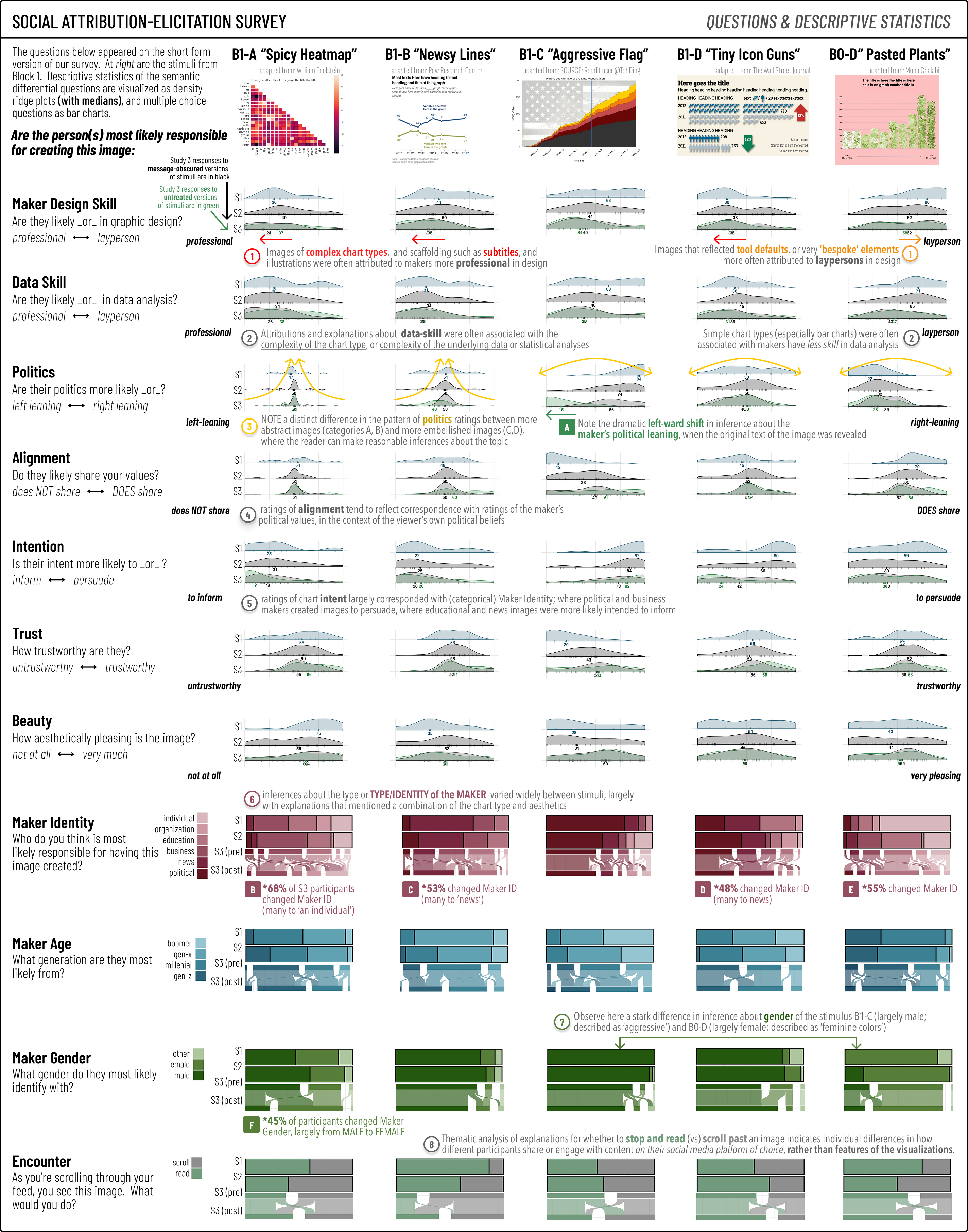}
  \caption{
    \textbf{Survey Questions Response Distributions for Block 1 Stimuli.} At \textit{left} find the short-form survey questions, and at \textit{right}, visualizations of the distribution of responses for each study (vertically stacked: S1, S2, S3) and each stimulus in Block 1 (as columns). 
  }
  \label{fig:results:descriptives}
  \vspace{-5mm}
\end{figure*}

If participants were generally unable to make attributions, or could only do so at the prompting of an interviewer, we would expect to see a high degree of short or nonsense text in the free-response data. In reviewing the explanations gathered in Studies 1 \& 2, 
we found respondents stated specific social attributions, similar in form to those reported 
in our interview study \cite{PAPER1}, 
as well as indications when they were not able to make attributions (i.e. answer the semantic differential questions with certainty.)
For example, in response to the stimulus \code{B1-B} (a dual-time series with simple colour palette), participant \code{S323} indicated 95\% confidence in their identification of the \textsc{Maker(Type)} as a news outlet, explaining, 
 \textit{``This looks like a newspaper like The New York Times." } However, there were also respondents who indicated reluctance to make attributions for certain \textsc{Actor(Properties).}
Participant \code{S328} indicated high confidence in their identification of the same stimulus as produced by a news outlet, but indicated 0\% confidence in their identification of \textsc{Maker(Gender)}, writing \textit{``I have no inclination one way or the other about the gender of the writer because that has nothing to do with [...] the source of the image.''} 
This increased our confidence that, while the free-response questions followed the semantic-differential scales and thus primed respondents’ attention toward social provenance, participants could nonetheless indicate (via the free-response) if they found the questions about social provenance difficult to answer.
Across both studies, free-response explanations frequently described inferences about \textsc{Maker(Type)} (e.g. news outlet, political party, company, etc.) based on a combination of chart type and ``editorial style''\cite{alebriEmbellishmentsRevisitedPerceptions2024} of the aesthetics. 
In contrast, participants gave fewer explanations of  \textsc{Maker(Gender)}, unless there was a specific feature (such as the pink background of stimulus \code{B0-D}, or ``aggressive colors'' of \code{B1-C}\textemdash(see \cref{fig:results:descriptives}) that strongly index a particular gender stereotype. 
This is consistent with the claim that an individual may not have socio-indexical associations between design features and every \textsc{Actor(Property)}, and further, that stimuli differ in their degree of ``indexical-salience'' between participants\cite{PAPER1}. In \autoref{fig:results:framework}—\textit{rightmost column} we illustrate the variety of social inferences explained by our participants in the free-response text, using the structure of our analytic framework for social provenance.

\subsubsection{Variance in Confidence}
\label{sec:results:replication:confidence}

\begin{figure}[h!]
  \vspace{-2mm}
  \centering 
    \includegraphics[width=1\columnwidth,alt={
    This figure  displays three rows of kernel density histograms, where each row corresponds to one confidence question from the study one and two surveys. The top row illustrates the distribution of responses to the maker-confidence question for each stimulus in block one. The visualizations indicate the distribution of confidence scores are a different shape for each stimulus, but do not differ between study one and two.  The same pattern is present in the second row illustrating the distribution of responses to the age confidence question. The final row displays distributions for the gender confidence question.  Here we see a difference in responses to two stimuli (B1-A and B1-B) between Study 1 (Tumblr sample) and Study 2 (Prolific sample). 
    }]{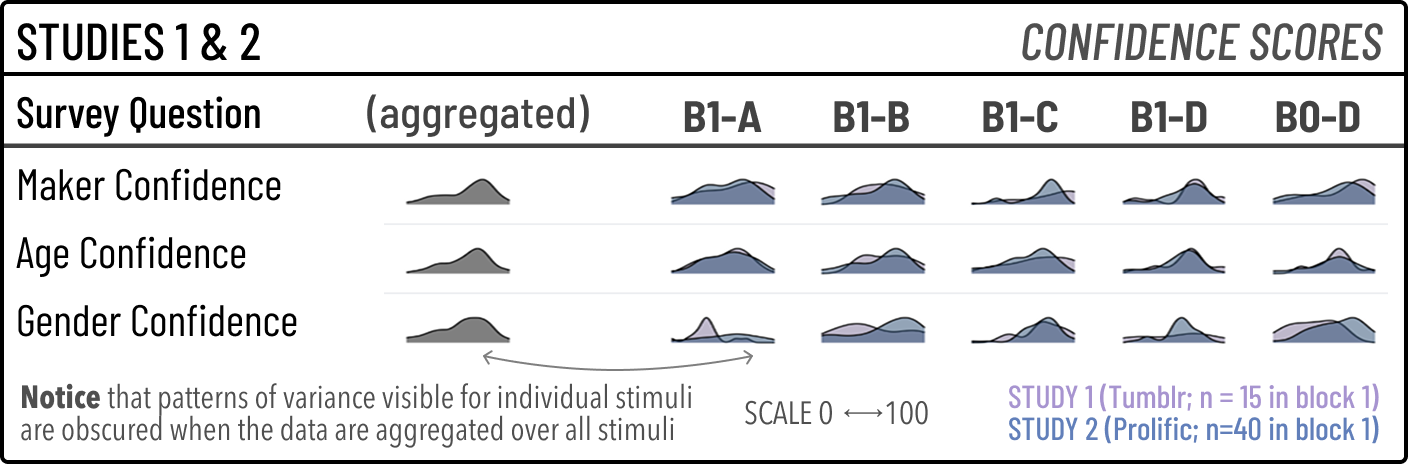}
  \caption{
    \textbf{Variance in Confidence Scores.} Responses to Studies 1\& 2 confidence questions illustrate how confidence varies across the stimuli. 
  }
  \label{fig:results:confidence}
  \vspace{-4mm}
\end{figure}

While participants can directly signal reluctance to answer questions to 
an interviewer, survey respondents cannot. 
In all surveys, participants were asked to indicate their confidence after answering the \textsc{Maker(Type), Maker(Age) and Maker(Gender)} multiple choice questions. 
If participants were generally unable to make attributions, or could only do so at the prompting of an interviewer, we would expect to see low variance in confidence scores between stimuli (with means clustered around 0 or 50\%)\textemdash indicating survey takers had consistently low confidence\textemdash\textit{or} uniform distributions for each stimulus\textemdash indicating a random response bias\cite{wetzelResponseBiases2016}. \autoref{fig:results:confidence} illustrates the distribution of each confidence question (for Block 1 stimuli) in Studies 1 \& 2.  Note that although the distribution of confidence ratings aggregated over all stimuli are similar, when broken out by stimulus, we see interesting patterns of variance. These patterns indicate that confidence in each kind of identification \textit{varied in response to different stimuli}\textemdash consistent with the claim that social attributions are made in response to the features of a particular visualization.

\subsubsection{Variance in Identifications \& Characterizations}
\label{sec:results:replication:semantic-differentials}

To explore how the features of
a visualization's design index social attributions, 
we must zoom in from our aggregated data to view responses to individual stimuli. 
\autoref{fig:results:descriptives} illustrates the distribution of responses for the ( n = 95 ) participants assigned to Stimulus Block 1 in all three studies. \textbf{These stimuli demonstrate a number of patterns salient in the other stimulus blocks, described in \autoref{fig:results:descriptives}-(annotations 1:8)}. 
\textit{Descriptive statistics and visualizations for each stimulus block are available in the \href{https://doi.org/10.17605/OSF.IO/23HYX}{supplemental materials}.}
If participants were generally unable to make attributions we would expect to see either: (1) a pattern of \textit{consistent response} in the semantic differential questions for each stimulus (e.g. uniform distributions for each stimulus, indicating random-answering), (2) stimulus-level clusters at 50\% (neutral response bias), or (3) stimulus-level bi-modal distributions (extreme response bias) \cite{wetzelResponseBiases2016}. Inspection of the stimulus-level distributions for each question (\autoref{fig:results:descriptives}) suggests this is not the case. To quantitatively verify that the patterns of variance observed do not result from these forms of response bias, we fit a linear mixed effects model on the semantic differential questions, predicting response \code{VALUE} by a linear combination of \code{STUDY} and interaction between \code{QUESTION} and \code{STIMULUS}. A model ANOVA indicates both significant main effects, and interaction term between \code{STIMULUS} and \code{QUESTION} ($F(240) = 15.5, p <0.001$)\textemdash indicating that participants' responses were influenced by both the Stimulus and Question being posed, which we would \textit{not} expect to find if participants responded following a pattern of neutral or extreme response bias \cite{wetzelResponseBiases2016}. Further, the main effect of \code{STUDY} was \textit{not significant} ($F(1) = 2.1, p=0.15$)\textemdash indicating that responses did not systematically vary as a function of the sample population (Study 1: Tumblr, Study 2: Prolific). Taken together, the patterns of variance in confidence and semantic differential questions, along with our inspection of free-response text across both studies,
suggest that:
(1) participants were in fact offering social attributions rather than nonsense responses, 
(2) that such attributions are not unique to users of Tumblr, and 
(3) the patterns of responding illustrate contours of the underlying phenomenon: inferences about the social provenance of a visualization. In the sections that follow, we combine responses from Studies 1 \& 2 for exploratory analysis.

\subsubsection{A Latent Structure of Social Attributions}
\label{sec:results:replication:latent-structure}

\begin{figure}[h]
  \centering 
    \includegraphics[width=1\columnwidth,
    alt={Figure 6 reports the factor loadings of an exploratory factor analysis on the Study 1 and 2 combined data.  The content of the three resultant factors is described in the manuscript. Additionally this table indicates that the MAKER-DESIGN  question loaded only on factor 2. The MAKER-DATA question loaded only on factor 3. The MAKER-POLITICS question loaded only on factor 1. The MAKER-TRUST question loaded on both factors 1 and 3. The MAKER-ALIGNMENT question loaded only on factor 1.  The CHART-BEAUTY question loaded on both factors one and two. Finally the MAKER-INTENT question loaded only on factor three.  The H-squared value is communality, and represents the different portion of each variable explained by the extracted factors, an indication of how well the factors account for the variance in the variable. The MAKER-DESIGN question has the highest communality at 99\%. The MAKER-POLITICS question had the lowest communality at 28\%. 
}]{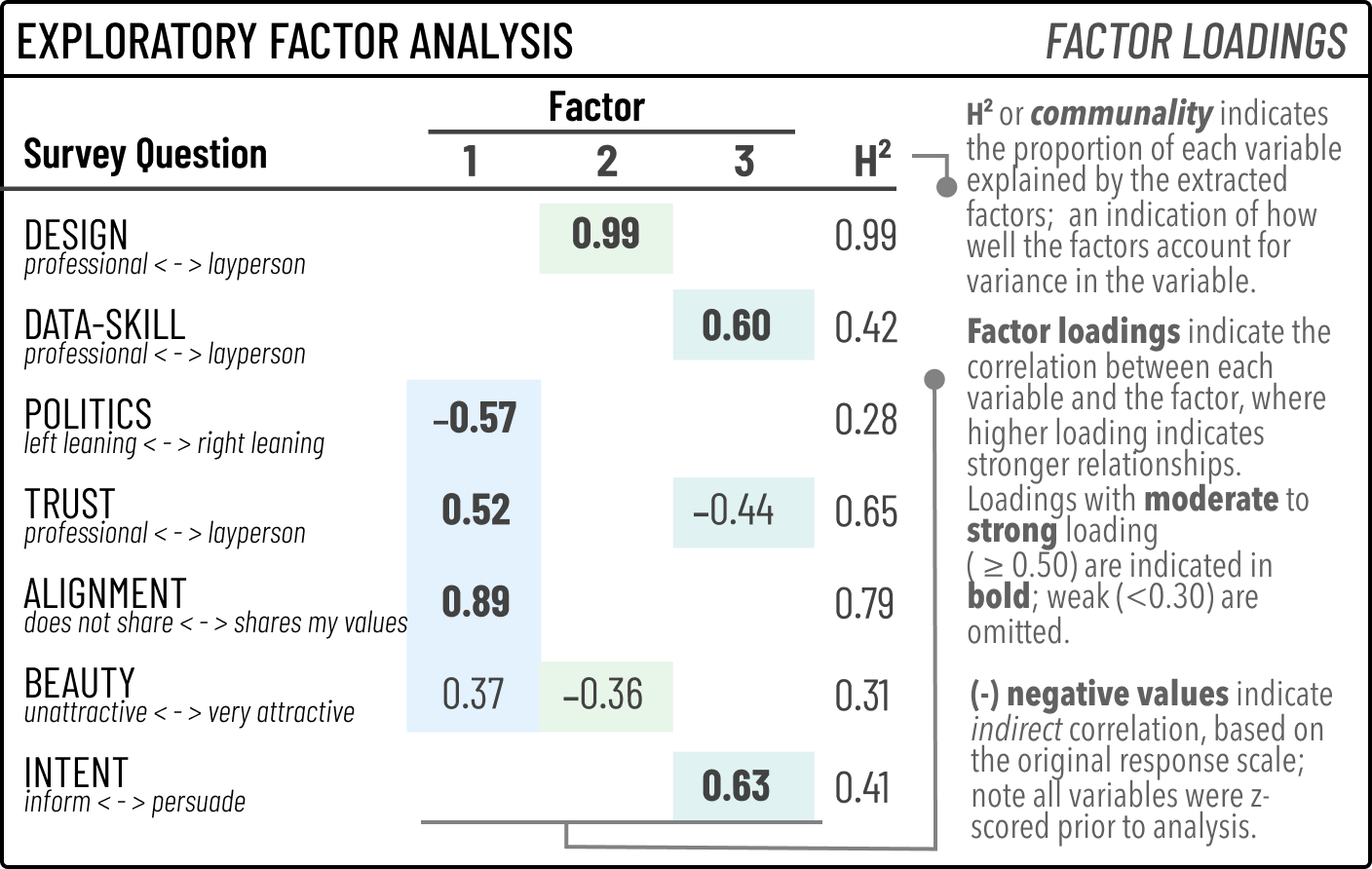}
  \caption{
    \textbf{Exploratory Factor Analysis.} Factor loadings for Studies 1 \& 2; observations of ( q = 7 ) semantic differential questions ( n = 318 )
  }
  \label{fig:s2:results:efa}
  \vspace{-5mm}
\end{figure}

To explore the structure of the survey data and identify potential latent constructs underlying the attributions measured via the 
semantic differential questions, we conducted an exploratory factor analysis\footnote{Like PCA, EFA is a dimensionality reduction technique; though EFA assumes variance in the observed data comes from unobserved \textit{latent constructs} (unobservable states/traits), and is used in psychometrics to determine how items on a measurement instrument are related to the psychological processes they measure; especially in the case of instruments with highly correlated questions.}. 
We first evaluated the suitability of the data by calculating the Kaiser-Meyer-Olkin measure ($KMO = 0.83$) (indicating a high level of sampling adequacy), noting Bartlett's test of sphericity was statistically significant ($\chi^2(21) = 2671, p < 0.001)$ (supporting the factorability of the correlation matrix). We used the maximum likelihood extraction method to determine the number of factors to retain (verified by a scree plot indicating a clear decrease in power after the third factor), and applied an \code{oblimin} rotation 
to enhance factor interpretability, 
appropriate for the high correlation between measures. The resulting factor loadings matrix is described in \autoref{fig:s2:results:efa}. The analysis identified three latent factors accounting for a cumulative 55\% of variance in the data:

\begin{enumerate}
    \item \textbf{A Trust/Alignment factor explains (23\%) of variance:}
    with attributions of the makers' political learning, trustworthiness, alignment with the viewer's values, and beauty of the image. 
    \item \textbf{A Design / Beauty factor explains (18\%) of variance:}  with the maker's skill in graphic design, and beauty of the image.   
    \item \textbf{A Data-Skill/Intent/Trust factor explains (14\%) of variance:} with ratings of maker trust, markers' intent (inform~$\leftrightarrow$~persuade) and skill in data analysis (professional~$\leftrightarrow$~layperson).  
\end{enumerate}

\noindent The high proportion of variance explained by the first Trust/Alignment factor indicates that these especially value-laden characterizations of the maker (including their political leaning and extent to which the maker shares the reader's values) are important in relation to trust. We address this relationship further via statistical modelling in (\cref{sec:results-trust}). 

\begin{figure*}[tb]
  \centering 
  \includegraphics[width=1\textwidth,
  alt={Figure 7 consists of two images reporting the results of the best fitting linear mixed effects model. The model is defined as predicting trust by a linear combination of an interaction between (beauty and intent) plus (alignment and intent) plus maker-data with a random intercept for participant. The first image is a line graph visualizing the model predicted outcomes for trust, at three values of alignment, beauty and intent: at -1, 0 and +1 standard deviations. The visualization illustrates the following main effects. The main effect of intent is such that informative intent is more trustworthy than persuasive intent. The main effect of beauty is such that more aesthetically pleasing images are more trustworthy. The main effect of the alignment is such that shared values are associated with more trust.  The following interaction effects are also illustrated.  Intentions to inform rather than persuade corresponded to decreasing the effect of beauty on trust. Persuasive intentions corresponded with increasing the effect of beauty on trust.  High alignment (shared values)  corresponded with decreasing the effect of intention on trust.  The second image consists of a table, reporting the standardized beta model estimates, confidence intervals, and P values for each term in the model.  The standardized beta for the beauty predictor is 0.11 with a confidence interval of [-0.072 to  +0.15], and p value less than 0.001.  The standardized beta for alignments is 0.46 with a conference interval from 0.43 to 0.50 with a p value less than 0.001. The standardized beta for intent is -0.23 with a confidence interval ranging from -0.26 to -0.19 and p value less than 0.001.  The standardized beta for maker-data is -0.17 with confidence interval ranging from -0.20 to -0.13 and p value less than 0.001. The standardized beta for the interaction term of beauty X intent is  0.05 with a confidence interval ranging from 0.022 to 0.09 and value less than 0.05. The standardized beta of the interaction term alignment by intent is 0.05 with a confidence interval ranging from 0.02 to 0.08 and p value less than 0.05. 
}]{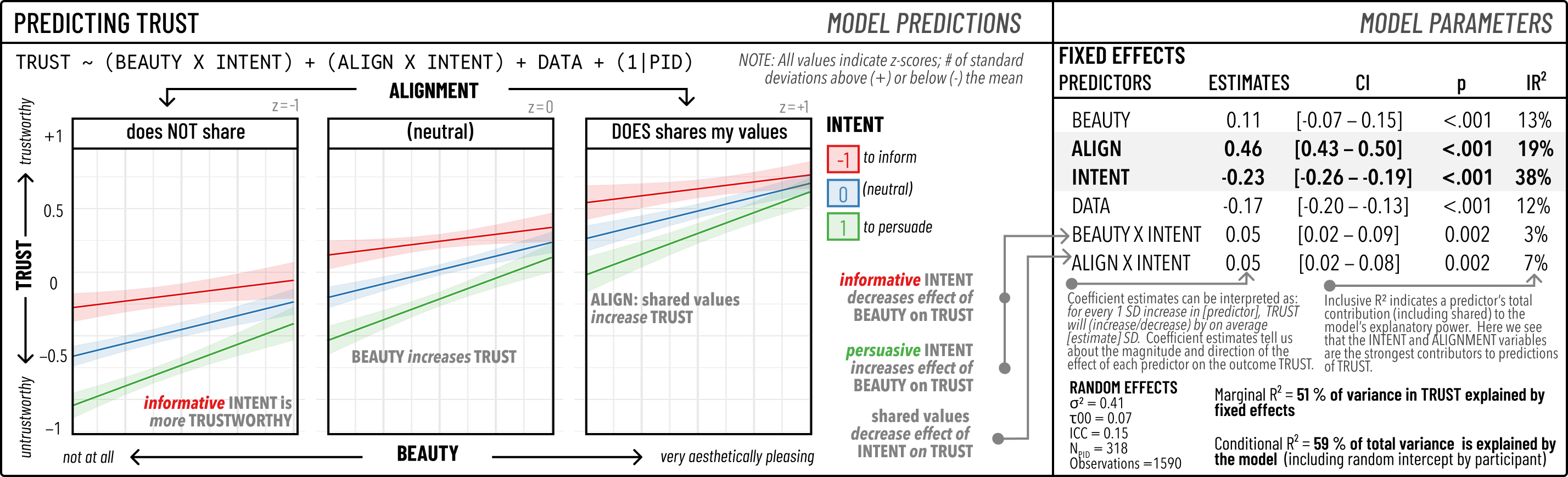}
  \caption{
  \textbf{Social Inferences Complicate the Relationship Between Beauty \& Trust.} 
  This linear mixed effects model explains $R^2(cond)=51\%$ of variance in \code{TRUST} by main effects of \code{BEAUTY}, \code{ALIGNMENT}, \code{INTENT} and \code{DATA-SKILL}, \& interactions (\code{INTENT} and \code{BEAUTY}) and (\code{INTENT} and \code{ALIGNMENT}).
  }
\label{fig:results:modeltrust}
\vspace{-6mm}
\end{figure*}

\color{black}
\subsection{RQ2: Predicting Trust: Inferences About Makers Matter}
\label{sec:results-trust}

Recent work in social psychology has extended a well-known effect from research on human faces\textemdash more attractive faces are deemed more trustworthy\textemdash to the case of visualization, offering crowd-sourced evidence that more beautiful visualizations are rated as more trustworthy \cite{linVisualizationAestheticsBias2021a}. 
However, participants in our interview study described
some graphs as being so well designed they were read as advertisements and thus untrustworthy, and alternatively, relatively `ugly' graphics as created by scientists and thus more trustworthy \cite{PAPER1}. Following this evidence, we hypothesize that the impressions a reader forms about the skills, values, and intentions of a visualization's maker likely moderate any relationship between trust and beauty. 
To explore this hypothesis, we compared a series of linear mixed effects models predicting \code{TRUST} by \code{BEAUTY}, as well as \code{ALIGNMENT} (does \textit{NOT} $\leftrightarrow$ DOES share my values), \code{INTENT} (inform $\leftrightarrow$persuade), and \code{DATA-SKILL} (professional $\leftrightarrow$layperson), with \code{PARTICIPANT} as random intercept.
We built models in a step-wise fashion deciding to keep successive predictors via $\chi^2$ and likelihood ratio tests. 
An initial model predicting \code{TRUST} by \code{BEAUTY} alone explained ($R^2{conditional}=22\%$) of variance in \code{TRUST}, with ($R^2{marginal}=12\%$) explained by the main effect of \code{BEAUTY}.   
By comparison, the best fitting model (shown in \autoref{fig:results:modeltrust}) explains ($R^2{conditional}=59\%$) of variance in \code{TRUST}, with ($R^2{marginal}=51\%$) explained by the following fixed effects:

\begin{itemize}[leftmargin=1em, itemindent=0.2em]
    
    \item A main effect of \code{BEAUTY} such that more \textit{attractive} graphs are rated as \textit{more} trustworthy. 
    \vspace{-1mm}
    
    \item A main effect of \code{ALIGNMENT} such that graphs from makers who \textit{share the viewer's values} are rated as \textit{more} trustworthy. 
    \vspace{-1mm}
    
    \item A main effect of \code{INTENT} such that graphs intended to \textit{inform} are rated as \textit{more} trustworthy than those intended to \textit{persuade}.
    \vspace{-1mm}
    
    \item A main effect of \code{DATA-SKILL} such that graphs from makers more \textit{professional} in data analysis are rated as \textit{more} trustworthy.
    \vspace{-1mm}
    
    \item An interaction of \code{BEAUTY} and \code{INTENT}, such that the effect of \code{BEAUTY} on \code{TRUST} is \textit{minimized} when the maker's intent is to  \textit{inform} rather than persuade \textit{(i.e. A graph intended to inform does not need to be as attractive as a graph intended to persuade in order to be assessed as trustworthy).} 
    \vspace{-1mm}
    
    \item An interaction of \code{ALIGNMENT} and \code{INTENT} such that the effect of 
    \code{INTENT} on \code{TRUST} is \textit{minimized} when the maker's values \textit{are aligned} with those of the viewer  \textit{(i.e. If I think the maker shares my values, their intent has less effect on my assessment of trust. If they do not share my values, then the trustworthiness of a graph intending to persuade is much lower than that of a graph intending to inform).} 
\end{itemize}

\noindent Taken together, these results demonstrate that in addition to a visualization's aesthetic appeal, a viewer's inferences about a maker's skills, values and intentions strongly influence trust. In our best fitting model the variable \code{INTENT} contributed to explaining ($IR^2 = 38\%$) of variance in trust, while \code{BEAUTY} contributed only ($IR^2 = 13\%$). To better understand how readers draw conclusions about a visualization's trustworthiness, we believe it is necessary to explore the communicative expectations of particular sociocultural groups, exploring differences in what design features index a visualization's communicative intent, and what makes an image `beautiful' in the eyes of a particular beholder.

\subsection{RQ3: Design Features \& Data Context}
\label{sec:results:study3-shift}

How do the combination of design features with topic \& takeaway messages of a visualization affect social inferences? In Studies 1 \& 2, participants viewed message-obscured versions of the stimuli 
in order to explore to what extent viewers make inferences about social provenance on the basis of the graph's design features alone, without reference to the messages of the underlying data. 
While complete obfuscation of data-topic is not possible when embellishments include iconic signs, we noted in the free-response data that even when viewers made reference to the presumed topic of the graph, the iconicity of the figural elements alone did not specify the perspective or `takeaway message'. For example, while the guns in \code{B1-D} and hospital beds in \code{B2-D} indexed the topics of violence and medical care respectively, they did not indicate if they were communicating `pro-gun' or `universal healthcare' messages. This provided the motivation for Study 3, where we used a repeated-measures design to ask participants the same set of questions about each image (in stimulus Block 1) twice: first, when viewing the message-obscured image, followed by the image with original text.  On the second viewing participants were also asked to explain how their answers did or did not change. 

To address our third research question we began by determining which attributions changed when viewed with the artifact's full text. To examine how \textit{characterizations} (semantic differentials) changed,
we evaluated a linear mixed effects model predicting continuous outcome \code{SHIFT} (change in answer between text obscured/unobscured) by a linear interaction between \code{QUESTION} and \code{STIMULUS}, with participant as a random intercept. As expected, an ANOVA indicated significant main effects of \code{QUESTION} and \code{STIMULUS} and significant interaction. 
Pairwise comparisons indicate that 
answers shifted the most for stimuli \code{B1-A} (a colorful heatmap), and \code{B1-C} (stacked area chart atop a US flag), and across all stimuli, the most labile questions were those about the maker's \code{POLITICS}, \code{TRUST}, and to what extent the maker's values \code{ALIGN} with the viewer's. We also evaluated a mixed effects logistic regression model predicting (binomial) \code{SHIFT} in answer to the \textit{identification} type questions (\code{MAKER TYPE, AGE, GENDER}) and similarly found significant main effects of both predictors, with post-hoc tests indicating participants most frequently changed their identification of the \code{MAKER TYPE}, and that of the stimuli in block 1, participants most frequently changed their identifications for stimulus \code{B1-A}. \textit{Significant effects are indicated in  \autoref{fig:results:descriptives}\textemdash(annotations A:F}). 

Inspecting the free response data helps us understand how data context influences social inferences. For many readers the initial identification of \code{B1-A} (a colourful heatmap) was an educational or scientific maker. However, when the text revealed the \textit{topic} of the graph was Spices, many changed the maker \code{TYPE} to \textit{an individual} and the \code{GENDER} to \textit{female}, while the \code{INTENT} shifted toward \textit{inform} and the graph became more \textit{trustworthy}, despite the individual maker shifting to more \textit{amateur} in \code{DATA} analysis.  Respondent \code{S47} at first explains, \textit{``I think its definitely intended to inform as there is no wording or descriptions to persuade. I think a business probably created this. Can't tell if man or woman or left vs right without more information.''} But upon viewing the full-text image, they write, \textit{``My answers changed when i saw the chart had to do with spices and is from allrecipes. This informs me an individual created [it]. I would guess most likely a woman as they cook more than males and probably search allrecipes more also.''} 

We also observed a dramatic shift in inferences about the maker of stimulus \code{B1-C} (stacked area chart atop a US flag).
In all studies the obscured version of this graph was overwhelmingly attributed to a right-leaning maker with persuasive intent, described as: \textit{``intentionally confrontational''}(\code{S359}),  \textit{``antagonistic''}(\code{S1479}), \textit{``pushing some narrative''}(\code{S304}). 
However the un-obscured text reveals the data concern types of corruption in the first administration of US-President Donald Trump (see \cref{fig:teaser}). With this text, readers' characterizations of the makers' \code{POLITICS} shifted dramatically to the \textit{left} with smaller but also significant shifts in alignment of values, but little change across the other attributions.  As \code{S85} explains, \textit{``I was definitely wrong about the political leaning [...] but I feel like most of my answers remained the same, based on its aesthetics.''} 
The striking aesthetic decisions made by \code{B1-C}'s maker and consequently strong (and consistent) inferences made by viewers indicate an effect with substantial implications for visualization design.  It is possible, through aesthetic design decisions to \textit{imply} that the identity and values of the maker \textit{align} with those of the target reader, even when that is not the case.  If it is true that humans prefer to engage with information that reinforces their existing world views, then representing conflicting data via aesthetics that lead the reader to infer the artifact was made \textit{by} and \textit{for} someone \textit{like them}, there is a pathway to engaging adversarial audiences. This effect is not unique to a particular political stance, as evidenced by this evocative response to an obscured stimulus \code{B1-B}: \textit{``That's a New York Times graph if I've ever seen one. It gave me a fight or flight response'' (\code{S324})}.

%% file: SECTIONS/06_discussion.tex
\vspace{-1.5mm}
\section{Discussion}
\label{sec:discussion}


In this paper we contribute a conceptual replication and extension of our recent work arguing that visualizations carry more than the semantic meaning of the data they encode; they also convey socio-indexical meaning (Morgenstern \& Fox et al.\cite{PAPER1}). 
We describe three attribution-elicitation studies demonstrating that: 
(1) social inferences can be elicited asynchronously via surveys; 
(2) such inferences are not unique to Tumblr \cite{PAPER1}, and can be elicited from a broader population sample;
(3) social inferences affect viewers' assessments of trust; and 
(4) social inferences are drawn from a combination of a visualization's design features, data-topic, and salient messages, in the context of the viewer's sociocultural identities. 
Taken together, we provide converging evidence for the \textit{socio-indexical function} of visualization. 
 When encountering visualizations, viewers not only read to extract insights about depicted data, but also make rich and nuanced inferences about an artifact's social origins and purpose in the world. 
We call for
broadening our perspective of how visualizations work\textemdash beyond the iconic and symbolic semiotic functions, to include \textit{indexicality}.
We argue this phenomenon opens a door to understanding situated behaviour with visualizations, such as how and why an individual might choose to engage or not engage with an artifact, 
and why artifacts might provoke adversarial readings. 
To catalyze future research 
we contribute an analytic framework useful for eliciting or analyzing discourse about social provenance, or as a reference for designers considering what social meaning their design choices might convey. 

\subsection{Limitations}
\label{discussion:limitations}

Although surveys are less subject to the demand characteristics imposed by live social interaction, they are nonetheless subject to social-desirability and other forms of response bias. This can limit a participant's willingness to express ideas that might be perceived as stereotypes. 
This presents a challenge to any explicit measure of socio-indexical phenomena.
Developing associations between formal/design features of communication and the subjective qualities and characteristics of those producing them involves social categorization, and invoking learned stereotypes associated with members of that social group. 
Acknowledging such stereotypes can be socially fraught. 
In order to mitigate such effects,
we recommend that future work explore \textit{implicit} measures of evaluative responses, 
drawing on a similar line of research in sociolinguistic cognition \cite{chevrotEditorsIntroductionReview2018}
in order to determine if social attributions are made spontaneously, or only when prompted.
Beyond response bias, the present studies are also limited in their ability to reveal systematic variance in social attributions that may exist \textit{between} sociocultural groups. By using a broad demographic sample 
our results likely reflect patterns in social attributions 
that are relatively consistent across populations. However, this sampling approach can mask indexical associations of crucial importance to developing targeted communication strategies for particular audiences. 
It may be beneficial for basic research to continue with broader samples to explore mechanisms 
or identify shared visualization ideologies.
However, we recommend that work seeking to develop design interventions take a more targeted approach.
That is, mapping the indexical fields for the target genre of communication and intended audience first through ethnographic or contextual inquiry. Most importantly, readers should be aware that the specific results of these studies are grounded in the sociocultural identities of participants sampled, the corpus of stimuli, and the situational context of engagement (i.e. a social media feed). If social inferences function in a way that is substantively similar to socio-indexical inferences in (natural) language, then a wealth of further work is required to describe: (1) the nature of visual language varieties, and (2) range of evaluative responses we may hold toward them. 

\subsection{Extending the Framework for Social Provenance}
\label{sub:discussion:framework}

The \hyperref[fig:results:framework]{analytic framework} we contribute 
is structured to aid thinking about \textsc{Inferences} composed of \textsc{Attributions} about \textsc{Properties} of \textsc{Actors}.
It reflects the structure of social provenance of visualizations in which we participate as researchers. 
However, we note that its \textit{content} 
(named Actors and Properties) is grounded in the design decisions of our studies, as well as the prior work which it extends\cite{PAPER1,kauerPublicLifeData2021a,hullmanContentContextCritique2015}. 
Research employing different populations, stimuli, and situational contexts 
may yield discourse revealing: (1) additional \textsc{Actors}, and 
(2) emphasizing inferences about different \textsc{Actor(Properties)}. 
For example, as we previously 
foregrounded interaction on social media \cite{PAPER1}, 
those interviews surfaced the \textsc{Maker—Animator}
as a salient \textsc{Actor}.
We would not necessarily expect this to be as salient in contexts where artifacts are not regularly shared by \textsc{Actors} uninvolved in their creation. 
Similarly, we expect populations with specialized expertise in visualization design (including technical communication) to yield more \textsc{properties} of \textsc{tools} with second-order inferences about the \textsc{Maker}'s competencies based on their tool use (such as programming languages \& libraries). 
We invite scholars studying engagement\textemdash especially in applied settings\textemdash to actively contribute to extending the framework, giving the community a more powerful tool for thinking about social provenance.

\subsection{Implications for Design: Encoding Social Provenance}
\label{discussion:design_implications}

Intended or not, we argue that every decision a designer makes, from chart type, to colour, framing text and modes of distribution, serve both to communicate desired insights about data, as well as impressions of the artifact's social history. Of course designers are not ignorant to the idea of understanding one's audience; this is central to design education and designerly ways of knowing \cite{parsonsDataVisualizationPractitioners2020a, parsonsDesignCognitionData2023a,DesignerlyWaysKnowing2006}. Nonetheless it seems that the degree of tuning (especially of aesthetics) to the expectation of one's audience that is prevalent in domains like graphic design, sits in tension with certain modern sensibilities prioritizing minimalism in 
visualization design \cite{cesalWhatAreData2021,schwabishWhyYourOrganization2021,Tufte1985,Tufte1990,Tufte1997, kostelnickVisualRhetoricData2007}. We believe this is a tension best resolved by the design community, but can be productively informed by adding a new dimension to empirically derived guidelines for visualization design that actively accounts for the role of socio-indexical inferences in how an artifact is received. An urgent example of the need for this collaboration comes from our exploration of trust. Our quantitative data indicate that social inferences can disrupt an otherwise straightforward hypothesis about beauty and trust: that the more \textit{aesthetically-appealing} a graphic is, the more trustworthy it will be. Our evidence demonstrates this relationship is more complicated.
Our data shows that assessments of beauty are moderated by a reader's inferences about both
the maker's intent and presumed alignment with their own values. 
Our survey respondents described particular images as \textit{so} `well-designed' they believed them to be advertisements, which in turn were not trustworthy, and not interacted with (e.g. "trying to sell me something $\rightarrow$ keep scrolling"). Alternatively, the most bland graphics using the default settings of basic word-processing programs were described by some as ``earnest'' and ``more objective''. These data suggest that designers and science communicators alike have a thin needle to thread with respect to communicating competency and professionalism, without being ``so slick'' as to trigger attributions of persuasive intent. This challenge is further complicated by the fact that what qualifies as informative versus persuasive differs by individual, as does an individual's openness to data and its
interpretation \cite{heEnthusiasticGroundedAvoidant2024}.

\subsection{\textbf{Implications for Research: From Graphical Perception \& Cognition to Socially-Situated Behaviour}}
\label{discussion:vispsych}

A tremendous volume of research has addressed the psychology of visualization, including: early-stage graphical perception
\cite{franconeriThreePerceptualTools2021, SzafirEnsemble2016,rensinkVisualizationStimulusDomain2021}, 
higher-order graph comprehension
\cite{Pinker1990,Shah2005,Fox_VisPsych_theoriesmodels_2023}, 
and subsequent judgement, reasoning and 
decision-making 
with extracted information 
\cite{PadillaDecisionMaking2018, kim_bayesian-assisted_2021, kale_visual_2021}.
Empirical research  
has yielded design guidelines for 
affording fast and accurate insights \textit{about data} \cite{tverskyCognitivePrinciplesGraphic1997,Hegarty2011,franconeri2021science}. 
What existing models of cognition struggle to account for, however, is how viewers’ social and situational contexts shape 
interaction with visualizations \cite{Fox_VisPsych_theoriesmodels_2023, FoxHollan_VisPsych_researchprogramme2023,Roth2003}. 
While a growing body of work explores individual differences \& sociocultural influences on \textit{reading data} from graphs \cite{xiongCurseKnowledgeVisual2020,markantCanDataVisualizations2022a, alebriEmbellishmentsRevisitedPerceptions2024,karduni_bayesian_2021,kim_bayesian-assisted_2021} we join colleagues studying \textit{engagement} more broadly \cite{peckDataPersonalAttitudes2019a, heEnthusiasticGroundedAvoidant2024, kennedyEngagingBigData2016,mahyarTaxonomyEvaluatingUser2015,kennedyFeelingNumbersEmotions2018} to document distinct forms of behaviour. \textbf{Social inferences are \textit{socio-indexical}, rather than \textit{semantico-referential}, graph ``readings''.}  We argue that to understand the broader context of engagement with visualizations 
outside the laboratory, 
we must document these socially-situated \textit{beyond-data} behaviours and determine their relationship to graphical perception and cognition. 
Although the descriptive evidence offered in our prior
interviews \cite{PAPER1} as well as the present surveys  
provide converging evidence for the \textit{existence} of socio-indexical inferences as a class of behaviour arising during interaction with visualizations, both methods involve direct elicitation, and 
are limited in their ability to illuminate the \textit{mechanisms} of this behaviour.
What kind of cognitive activities lead to social inferences, and how is this activity related to graphical perception, comprehension, and decision-making? 
From the present work, we cannot describe the temporal dynamics 
of social inferences, nor predict if they occur spontaneously, or 
only upon elicitation.  
We suggest a logical next step 
is to join researchers at the intersection of sociolinguistics and cognitive science studying the mechanisms of language attitudes\cite{chevrotEditorsIntroductionReview2018}.
This would involve
adapting implicit measures of social attribution (such as sociolinguistic guise\cite{dragojevicVerbalGuiseTechnique2022a,loureiro-rodriguezMatchedGuiseTechnique2022a} and implicit association tests \cite{campbell-kiblerImplicitAssociationTest2012a, campbell-kiblerSociolinguisticsPerception2010a}), and other experimental paradigms. 
Such measures could 
reveal: the timescale of social inferences, 
what stages of cognitive information processing they 
affect,
and the consequences of inferences not explicitly expressed. 
If social inferences are made automatically, or in the timescale of object-recognition and attention
allocation, this has profound implications for 
what we know about graphical perception and comprehension. If social inferences are more volitional or occur on a longer timescale, they are nonetheless critical to understanding how and why people choose to interact with visualizations in particular ways, and how inferences about makers may bias our decisions and dispositions toward their encoded data.